\documentclass[twocolumn,prl,showpacs,floatfix]{revtex4}

\usepackage{subfigure}
\usepackage{graphicx}

\begin{document}

\preprint{DRAFT}
\date{February 14, 2017}

\title{Duration of classicality in highly 
degenerate interacting Bosonic systems}

\author{Pierre Sikivie and Elisa M. Todarello}

\affiliation{Department of Physics, University of Florida, 
Gainesville, FL 32611, USA}

\begin{abstract}

We study sets of oscillators that have high quantum occupancy and 
that interact by exchanging quanta.   It is shown by analytical 
arguments and numerical simulation that such systems obey 
classical equations of motion only on time scales of order their 
relaxation time $\tau$ and not longer than that.  The results
are relevant to the cosmology of axions and axion-like particles.

\end{abstract}

\maketitle

The question under consideration here is: on what time scale 
do highly degenerate, interacting quantum oscillators obey 
classical equations of motion?  Consider the broad class of 
systems that have a Hamiltonian of the form
\begin{equation}
H = \sum_j \omega_j a_j^\dagger a_j + 
{1 \over 4} \sum_{jkln} \Lambda_{jk}^{ln} 
a_j^\dagger a_k^\dagger a_l a_n
\label{Ham}
\end{equation} 
where the $a_j$ and $a_j^\dagger$ are annihilation and 
creation operators satisfying canonical equal-time commutation
relations.  ${\cal N}_j = a_j^\dagger a_j$ is the number of
quanta in oscillator $j$.  For the sake of definiteness, we
have restricted ourselves in Eq.~(\ref{Ham}) to systems in 
which the total number of quanta $N = \sum_j {\cal N}_j$ 
is conserved.   The system states are given by linear 
combinations
\begin{equation}
|\Psi(t)\rangle = \sum_{\{{\cal N}_j\}} 
c(\{{\cal N}_j\}, t)~|\{{\cal N}_j\}\rangle
\label{wvfct}
\end{equation}
of eigenstates $|\{{\cal N}_j\}\rangle$ of the ${\cal N}_j$ for arbitrary 
distributions $\{{\cal N}_j\} = ({\cal N}_1, {\cal N}_2, {\cal N}_3, ...)$
of the quanta over the oscillators.  In the Heisenberg picture, where the 
time-dependence of the state vectors has been removed, the annihilation 
operators $a_j(t)$ satisfy the equations of motion
\begin{equation}
i \dot{a}_j = [a_j, H] = \omega_j a_j +
{1 \over 2} \sum_{kln} \Lambda_{jk}^{ln} a_k^\dagger a_l a_n~~\ .
\label{Ehren}
\end{equation}
The classical description of the system is obtained by replacing 
the $a_j(t)$ with c-numbers $A_j(t)$.  They satisfy
\begin{equation}
i \dot{A}_j = \omega_j A_j + 
{1 \over 2} \sum_{kln} \Lambda_{jk}^{ln} A_k^* A_l A_n~~\ .
\label{ceom}
\end{equation}
The quantum description always requires vastly more information than 
the classical one.   To be specific, if the number of oscillators 
is $M$ and the number of quanta $N$, the classical state is given by 
$2M-1$ real numbers, whereas the quantum state is given by  
\begin{equation}
D = {(N + M -1)! \over N! (M - 1)!} - 1
\label{Hildim}
\end{equation}
complex numbers.   For example, if $N = 100$ and $M = 10$, 
$D = 4.26 \cdot 10^{12}$.  $D$ increases extremely fast 
with increasing $N$ and $M$.  Clearly a huge simplification 
occurs if the system obeys classical equations of motion.  The 
question is:  when is this approximation valid?

The question is particularly relevant to axion cosmology 
\cite{axdm,CABEC,axtherm,Saikawa,Berges,Guth}.  The number 
of axions inside a co-moving volume of size (1 Mpc)$^3$ today 
is $N \simeq 4 \cdot 10^{81}$, assuming all the dark matter 
is axions and the axion mass is $10^{-5}$ eV.  Before structure
formation, their momentum dispersion is at most of order 
$\delta p \sim {1 \over t_1}{a(t_1) \over a(t)}$ where 
$t_1 \sim 10^{-7}$ sec is the age of the universe when the 
axion mass effectively turns on, and $a(t)$ is the 
cosmological scale factor.  Their quantum degeneracy, i.e. the 
average occupation number of those states that the axions occupy, 
is thus at least of order ${\cal N} \sim 10^{61}$ \cite{CABEC}.   
Almost all discussion of the cosmology of axions \cite{axdm,
Guth,Braaten} or axion-like \cite{ALPs} particles assumes that 
the axion fluid obeys classical field equations.   However, it 
was shown in refs. \cite{CABEC,axtherm} that the axion fluid 
thermalizes on a time scale shorter than the age of the universe 
after the photon temperature has dropped below approximately 500 eV.  
When the axion fluid thermalizes, it satisfies all conditions for 
Bose-Einstein condensation and this should therefore be the expected 
outcome on theoretical grounds.  Furthermore it was shown \cite{case} 
that Bose-Einstein condensation of cold dark matter axions explains 
precisely and in all respects the observational evidence for caustic 
rings of dark matter in disk galaxies.  The evidence is summarized 
in ref. \cite{Duffy}.  Bose-Einstein condensation is a quantum effect.  
The argument that cold dark matter axions form a Bose-Einstein 
condensate was questioned \cite{Guth} in part on the belief that 
the cosmic axion fluid satisfies classical field equations as a 
result of its extremely high degeneracy.  This belief is also 
implicit in the many other discussions of dark matter axions, or 
axion-like particles, which describe the axion fluid by classical 
field equations \cite{ALPs}.  So, we want to ask: is it true that 
highly degenerate Bosonic systems obey classical equations of motion 
merely because they are highly degenerate?  And, if they obey classical 
field equations of motion for a while but not forever, what is the 
time scale over which classical equations of motion are obeyed?

When the interactions among the oscillators are turned off, i.e. 
when the $\Lambda_{jk}^{ln} = 0$, and the degeneracy ${\cal N}$ is 
high, a classical description is in fact correct, and accurate to 
order $1/{\cal N}$.  Indeed Eqs.~(\ref{Ehren}) and (\ref{ceom}) are 
linear in that case and admit solutions that have identical time 
dependence.  If the expected values 
$\langle{\cal N}_j\rangle \equiv \langle\Psi(t)|{\cal N}_j|\Psi(t)\rangle$ 
and their classical analogues $N_j = A_j^*(t) A_j(t)$ are equal 
initially, they remain equal ever after.  In spite of its
apparent ``triviality", the non-interacting case describes a large 
number of interesting phenomena where the system has a non-trivial 
evolution either because the initial state is a linear superposition 
of different eigenmodes (e.g. the beating of a double pendulum) or 
because the oscillation frequencies of the oscillators are time-dependent 
(e.g. parametric resonance).  Such phenomena are described by classical 
physics when ${\cal N}$ is large.  The production of cold axions by vacuum 
realignment in the early universe is a case in point. Because the effect is 
due to the time dependence of the axion mass and interactions do not play 
an important role, a classical physics calculation produces a correct estimate 
of the axion cosmological energy density from vacuum realignment \cite{axdm}.  
Perhaps the successes of classical physics when $\Lambda_{jk}^{ln} = 0$ and 
${\cal N} \rightarrow \infty$ has led to a widely held belief that classical 
physics also gives a good description when $\Lambda_{jk}^{ln} \neq 0$ and 
${\cal N} \rightarrow \infty$.

When $\Lambda_{jk}^{ln} \neq 0$, the $\langle{\cal N}_j\rangle$ are 
time-dependent because quanta jump between oscillators in pairs: one 
quantum jumps from oscillator $l$ to oscillator $j$ while another quantum 
jumps from $n$ to $k$.  The classical $N_j(t)$ are also time-dependent 
when $\Lambda_{jk}^{ln} \neq 0$.  The question here is whether the 
time dependence is the same.  Assuming the initial state is far 
from equilibrium, there exists a time scale $\tau$ over which the 
distribution of the quanta over the oscillators changes completely, 
i.e. each $\langle{\cal N}_j\rangle$ changes by order 100\%.  We call 
$\tau$ the relaxation time and $\Gamma = 1/\tau$ the relaxation rate.
If the system is stable, it will move toward thermal equilibrium 
on a time scale of order $\tau$.  If the system is unstable, it 
will also move towards thermal equilibrium on a time scale of 
order $\tau$ provided the time scale of instability is long 
compared to $\tau$.  

There is a simple a-priori reason to expect the quantum and 
classical descriptions to deviate from each other on a time 
scale of order $\tau$.  Indeed, the quantum description has 
the system move towards a Bose-Einstein distribution whereas 
the classical description has the system move towards a Boltzmann 
distribution. This argument is compelling but perhaps not precise 
enough to give us an estimate of the time scale of classicality.  
It allows the classical description to be valid, for example, on a 
time scale of order $\tau \log({\cal N})$.  For the systems that 
we are familiar with in the laboratory, mainly superfluid $^4$He 
and dilute ultra-cold atoms, the quantum degeneracy is not much 
larger than one.   So we have no compelling guidance from experiment 
to tell us about the behavior of systems with huge degeneracy 
such as the cosmic axion fluid with ${\cal N} \sim 10^{61}$.  

To gain insight, consider the evolution equations for the 
occupation numbers.  There are two cases to consider depending
whether $\Gamma < \delta \omega$, where $\delta \omega$ is the 
energy dispersion, or $\Gamma > \delta \omega$.  In the first
case, called the particle kinetic regime, we have  
\begin{eqnarray}
\dot{\cal N}_j = \sum_{kln} |\Lambda_{jk}^{ln}|^2
\pi \delta(\omega_j + \omega_k - \omega_l - \omega_n)\cdot~~~&~&
\nonumber\\
\cdot[({\cal N}_j+1)({\cal N}_k+1) {\cal N}_l {\cal N}_n 
- {\cal N}_j {\cal N}_k ({\cal N}_l+1) ({\cal N}_n+1)]&~&
\label{pkq}
\end{eqnarray}
for the operators ${\cal N}_j(t)$ in the Heisenberg picture 
\cite{axtherm}, and 
\begin{eqnarray}
\dot{N}_j &=& \sum_{kln} |\Lambda_{jk}^{ln}|^2
\pi \delta(\omega_j + \omega_k - \omega_l - \omega_n)\cdot
\nonumber\\
&\cdot&[(N_k + N_j) N_l N_n - N_k N_j (N_l + N_n)]
\label{pkc}
\end{eqnarray}
for the c-numbers $N_j(t)$ \cite{premature}.  In the second 
case, called the condensed regime, we have instead \cite{axtherm}
\begin{equation}
\dot{\cal N}_j = {i \over 2} \sum_{kln} 
(\Lambda_{ln}^{jk} a_l^\dagger a_n^\dagger a_k a_j 
- \Lambda_{jk}^{ln} a_j^\dagger a_k^\dagger a_l a_n)
\label{crq}
\end{equation}
and 
\begin{equation}
\dot{N}_j = {i \over 2} \sum_{kln}
(\Lambda_{ln}^{jk} A_l^* A_n^* A_j A_k 
- \Lambda_{jk}^{ln} A_j^* A_k^* A_l A_n)~~\ .
\label{crc}
\end{equation}
For a fluid of interacting particles, such as the cosmic axion 
fluid, the oscillators in Eq.~(\ref{Ham}) are labeled by the 
particle momenta $\vec{p} = {2 \pi \over L} (n_1, n_2, n_3)$ 
where the $n_r$ ($r$ = 1,2,3) are integers and $L$ is the 
linear size of a large cubic volume $V = L^3$ in which the 
associated quantum field satisfies periodic boundary conditions.  
The oscillator frequencies are $\omega_{\vec p} = {p^2 \over 2m}$ 
in the non-relativistic limit.  In the case of cosmic axions, the 
relevant interactions are $\lambda \phi^4$ and gravitational, for 
which the couplings are respectively
\begin{equation}
\Lambda_{\lambda~\vec{p}_1,\vec{p}_2}^{\vec{p}_3,\vec{p}_4}
= {\lambda \over 4 m^2 V} 
\delta_{\vec{p}_1+\vec{p}_2,\vec{p}_3+\vec{p}_4}
\label{lamlam}
\end{equation}
and
\begin{equation}
\Lambda_{g~\vec{p}_1,\vec{p}_2}^{\vec{p}_3,\vec{p}_4}
= - {4 \pi G m^2 \over V}
({1 \over |\vec{p}_1 - \vec{p}_3|^2} + 
{1 \over |\vec{p}_1 - \vec{p}_4|^2})
\delta_{\vec{p}_1+\vec{p}_2,\vec{p}_3+\vec{p}_4}~~\ .
\label{glam}
\end{equation} 
In the particle kinetic regime, Eqs.~(\ref{pkq}) and 
(\ref{pkc}) imply relaxation rates of order
\begin{equation}
\Gamma_{\rm pk} \sim n \sigma \delta v {\cal N}
\label{relaxpk}
\end{equation}
where $n$ is the physical space density, $\delta v$ is 
the velocity dispersion, and $\sigma$ is the appropriate 
cross-section. For $\lambda \phi^4$ interactions, 
$\sigma_\lambda = {\lambda^2 \over 64 \pi m^2}$.  For
gravity, the appropriate cross-section is that for large angle 
scattering,  $\sigma_g \sim {4 G^2 m^2 \over (\delta v)^4}$, 
since forward scattering does not contribute to relaxation.
In the condensed regime, Eqs.~(\ref{crq}) and Eqs.~(\ref{crc})
imply relaxation rates of order
\begin{equation}
\Gamma_{{\rm cr},\lambda} \sim {n \lambda \over 4 m^2}~~~{\rm and}~~~
\Gamma_{{\rm cr},g} \sim {4 \pi G n \over (\delta v)^2}
\label{relaxcr}
\end{equation}
respectively.  The relaxation rate estimates appear very 
different in the two regimes.  However they are related 
by $\Gamma_{\rm pk} \sim (\Gamma_{\rm cr})^2/\delta \omega$
so that they agree with one another at the inter-regime 
boundary where $\Gamma = \delta \omega$.  Axion dark matter
was found \cite{CABEC,axtherm} to thermalize in the condensed 
regime by their gravitational self-interactions when the 
photon temperture is of order 500 eV.  

Eqs.~(\ref{pkq}) and (\ref{crq}) for quantum evolution closely 
resemble their classical counterparts, Eqs. (\ref{pkc}) and 
(\ref{crc}). However, let us point out two significant differences 
between Eqs.~(\ref{pkq}) and (\ref{pkc}).  Similar differences
exist between Eqs.~(\ref{crq}) and (\ref{crc}).  The first and, 
as it will turn out, most important difference is that Eq.~(\ref{pkq}) 
is an operator equation whereas Eq.~(\ref{pkc}) is a c-number equation.  
The second difference is that the expression in brackets in Eq.~(\ref{pkq}) 
has terms ${\cal N}_l {\cal N}_n - {\cal N}_j {\cal N}_k$ that have no 
analogues in Eq.~(\ref{pkc}).   Indeed, if one attempts to derive 
Eq.~(\ref{pkc}) from (\ref{pkq}) by taking the quantum expectation value 
on both sides of Eq.~(\ref{pkq}) and identifying 
$\langle{\cal N}_j(t)\rangle$ with $N_j(t)$, 
one encounters two difficulties.  The first is that 
$\langle{\cal N}_j {\cal N}_k {\cal N}_l\rangle 
\neq \langle{\cal N}_j\rangle \langle{\cal N}_k\rangle \langle{\cal N}_l\rangle$.  
The second is that the expressions in brackets in the two equations are 
different even after replacing ${\cal N}_j$ by $N_j$.   

There are specific cases where Eqs.~(\ref{pkq}) and (\ref{pkc}) 
make dramatically different predictions because of the quadratic 
terms in Eq.~(\ref{pkq}) that have no analogues in Eq.~(\ref{pkc}).
For example consider the initial momentum distribution 
\begin{equation}
N_{\vec p} = \sum_{j=1}^J N_j \delta_{\vec{p},\vec{p}_j}
\label{example}
\end{equation}
with $N_j \neq 0$ for a set of momenta $\vec{p}_j$ ($j = 1, 2, ..., J)$ 
such that the process $\vec{p} + \vec{p}^{~\prime} \rightarrow 
\vec{p}^{~\prime\prime} + \vec{p}^{`\prime\prime\prime}$ violates 
energy-momentum conservation for any $\vec{p}$, $\vec{p}^{~\prime}$ 
and $\vec{p}^{~\prime\prime}$ belonging to the set, and arbitrary
$\vec{p}^{~\prime\prime\prime}$.  In other words, in this 
configuration any scattering allowed by energy-momentum
conservation is into final states that are both initially empty. 
This momentum distribution has a time-dependent evolution according 
to Eq.~(\ref{pkq}), whereas it is time-independent according to 
Eq.~(\ref{pkc}).  Indeed the process 
$\vec{p} + \vec{p}^{~\prime} \rightarrow
\vec{p}^{~\prime\prime} + \vec{p}^{~\prime\prime\prime}$
always occurs in the quantum theory when the initial modes
are occupied (${\cal N}_{\vec p} \neq 0$ and 
${\cal N}_{{\vec p}^{~\prime}} \neq 0$), whereas it occurs in 
the classical theory only if in addition one of the final 
modes is occupied ($N_{\vec{p}^{~\prime\prime}} \neq 0$ or 
$N_{\vec{p}^{~\prime\prime\prime}} \neq 0$).  As a particular case, 
two monochromatic particle beams do not scatter in the classical 
theory unless one of the final states allowed by energy-momentum 
conservation is already occupied, whereas two such beams always 
scatter in the quantum theory. The momentum distribution of 
Eq.~(\ref{example}) is not generic but even for generic initial 
momentum distributions the $\langle{\cal N}_j\rangle$ will deviate 
from the $N_j$ because of the extra terms in the brackets of 
Eq.~(\ref{pkq}) that have no analogues in Eq.~(\ref{pkc}).  
After a time of order $\tau$, the resulting difference 
$\langle{\cal N}_j\rangle - N_j$ will be ${\cal O}(1)$ and 
then grow exponentially fast as the quantum evolution leads to a 
Bose-Einstein distribution whereas the classical evolution 
leads to a Boltzmann distribution.  So the classical evolution 
will certainly deviate from the quantum evolution by order 100\% 
after a time of order $\tau \log {\cal N}$.   Actually, as we 
now show, the classical and quantum evolutions deviate from one 
another much faster than that because Eq.~(\ref{pkq}) is an 
operator equation whereas Eq.~(\ref{pkc}) is a c-number equation.

Only in eigenstates $|\{{\cal N}_j\}\rangle$ of the 
occupation numbers is $\langle{\cal N}_j {\cal N}_k {\cal N}_l\rangle
= \langle{\cal N}_j\rangle \langle{\cal N}_k\rangle \langle{\cal N}_l\rangle$.  
Generally speaking, it is exceedingly unlikely at any given moment 
that the quantum state is an eigenstate of the occupation 
numbers.  Even if it happens to be in such a state, quantum 
evolution will soon, as a result of interactions, cause it 
to become a linear superposition of many different $|\{{\cal N}_j\}\rangle$.  
In contrast, the classical state is always an eigenstate of the $N_j$.  
To investigate the implications of this difference, we carried out 
numerical simulations of five oscillators in the condensed regime.  
The toy system we use was first described in ref. \cite{axtherm} 
and shown there to thermalize on the expected time scale $\tau$.
Its Hamiltonian has the form given in Eq.~(\ref{Ham}) with  
$\omega_j = j \omega_1$ ($j$ = 1, 2, 3, 4, 5) and $\Lambda_{jk}^{ln} 
= 0$ unless $j + k = l + n$.  Non-zero values are given to 
$\Lambda_{14}^{23}$, $\Lambda_{15}^{24}$, $\Lambda_{25}^{34}$, 
$\Lambda_{22}^{13}$, $\Lambda_{33}^{24}$, $\Lambda_{33}^{15}$ and 
$\Lambda_{44}^{35}$, and their conjugates $\Lambda_{jk}^{ln} = 
\Lambda_{ln}^{jk~*}$.  We numerically integrate the Schr\"odinger 
equation 
\begin{equation}
i \partial_t |\Psi(t)\rangle = H |\Psi(t)\rangle
\label{Schrod}
\end{equation}
for this model starting from an initial state which is an eigenstate of the 
occupation numbers, calculate the expectation values $\langle{\cal N}_j\rangle(t)$ 
and compare with the classical evolution $N_j(t)$ obtained by numerically 
integrating Eqs.~(\ref{ceom}).   A large number of initial conditions were
simulated.  We find in all cases that the classical evolution deviates 
from the quantum evolution on a time scale which is short compared to 
$\tau$.  

\begin{figure}   
\includegraphics[width=0.95\columnwidth]{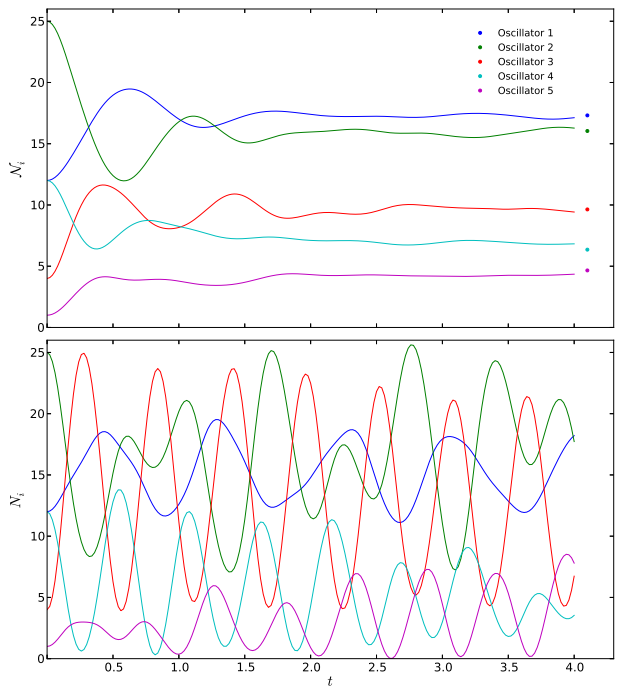}
\caption{Quantum (top) and classical (bottom) time evolution 
of the occupation numbers in the toy system described in the 
text for the initial state $|12,25,4,12,1\rangle$.  The dots 
on the right in the top panel indicate the thermal averages 
in the quantum case.  The quantum system approaches the thermal 
averages on the expected time scale.   The classical system does 
not equilibrate.  The classical evolution tracks the quantum 
evolution only very briefly.}   
\label{cq}   
\end{figure}

The top panel of Fig. 1 shows the quantum evolution of the initial state 
$|{\cal N}_1, {\cal N}_2, ..., {\cal N}_5\rangle =  |12, 25, 4, 12, 1\rangle$ 
as an example.  The figure shows that the expected values $\langle{\cal N}_j\rangle$ 
move towards the thermal averages on the expected time scale $\tau$, 
which is of order one given the coupling strengths $\Lambda_{jk}^{ln}$ 
in the simulation \cite{axtherm}.  The quantum thermal averages are 
computed by giving equal probability to each system state consistent 
with the total number of quanta and the total energy in the initial 
state.  They are shown by the dots on the right side of Fig. 1.  The 
bottom panel of Fig. 1 shows the classical evolution of the initial 
state $(A_1, A_2, ..., A_5) = 
(\sqrt{12}, \sqrt{25}, \sqrt{4}, \sqrt{12}, \sqrt{1})$, in which 
the $N_j$ and their time derivatives $\dot{N}_j$ have the same 
initial values as their quantum analogues in the top panel.  Fig. 1 
shows that the classical evolution tracks the quantum evolution only 
for a short time compared to $\tau$.  Fig. 1 also shows that the 
classical oscillators do not approach thermal equilibrium on the 
time scale $\tau$.  If the simulation is prolonged, one finds that 
the classical oscillators do not thermalize even after a very long 
time.  This phenomenon was first noted by Fermi, Pasta and Ulam in 
1955 and has been studied by many authors since \cite{Fermi}.

\begin{figure}   
\includegraphics[width=\columnwidth]{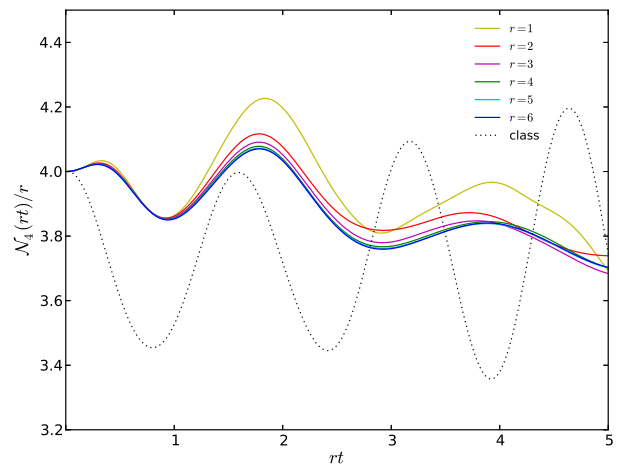}
\caption{Quantum evolution of the occupation number of the 
fourth oscillator (solid lines) of the toy system when the 
initial occupation numbers are $r\times$(6,4,5,4,3) with 
$r = 1, 2, ... 6$. Both axes are rescaled according to the scale
invariance of the classical equations of motion. The classical 
evolution is shown by the dotted line.  The figure shows that 
when $r$ is increased, the quantum evolution approaches a limit 
which differs from the classical evolution.}
\label{resc}   
\end{figure}

Our concern is whether the classical evolution is a good approximation
to the quantum evolution.   It is not for the initial condition 
described in the previous paragraph, nor for all the other initial 
conditions that we simulated.  One may wonder whether this due to 
the occupation numbers being too small.  To test this we did a 
series of simulations in which all the occupation numbers are 
scaled up by a common factor $r$.  The classical evolution remains 
unchanged under such a rescaling provided time $t$ is rescaled by 
$1/r$.  The quantum evolution is not rescaling invariant for small 
$r$ but is found in our simulations to approach a rescaling invariant 
limit when $r$ is increased, as shown in Fig. 2.  For the largest system 
simulated (r=6), the occupation numbers range from 18 to 36. Performing 
much larger simulations is prohibitively expensive. However, the convergence 
of the quantum behaviour for increasing r leads us to believe that a further 
increase of the occupation numbers would not produce any relevant changes.  
Thus we find that the quantum evolution is different from the classical 
evolution in the large ${\cal N}$ limit.

One may also ask whether the classical evolution equations give 
a good approximation to quantum evolution if the initial state is 
a coherent state, i.e. a state of minimum uncertainty in the $a_j$.
We tested this and found that it does not.  Fig. 3 shows three different 
evolutions of the initial state (0,12,16,0,0): i) the classical evolution, 
ii) the quantum evolution, and  iii) the quantum evolution of the 
corresponding coherent state.  Evolutions ii) and iii) are similar 
and different from evolution i).   

We conclude that highly degenerate interacting Bosonic systems
obey classical equations of motions only on time scales at most 
of order the relaxation time scale $\tau$.  Our simulations had 
only five oscillators but there is no reason to think that the 
classical description fares any better when the number of 
oscillators is increased.  Our result is relevant to the cosmology 
of dark matter axions and axion-like particles because their 
relaxation rate by gravitational self-interactions becomes, 
at some point, shorter than the evolution rate of the universe 
\cite{CABEC,axtherm}. When this happens, the commonly made assumption 
that the axion fluid obeys classical field equations is unjustified.  
Classical field equations are still valid, of course, as a description 
of stable or metastable objects in the axion fluid such as flat domain 
walls \cite{axwall}, straight strings \cite{axstring} and Bose stars 
\cite{Tkachev},\cite{Guth},\cite{Braaten},\cite{Davidson} since 
thermalization plays no role for them.

\begin{figure} 
\includegraphics[width=\columnwidth]{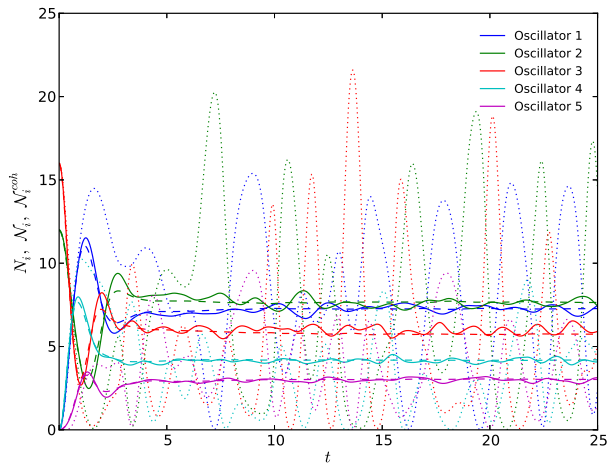} 
\caption{Quantum evolution (solid lines) and classical evolution 
(dotted) of the occupation numbers for the initial state (0,12,16,0,0).
The dashed lines show the quantum evolution of the corresponding 
coherent state.}
\label{coh} 
\end{figure}

We thank Edward Witten, Charles Thorn, Joerg Jaeckel, Adam Christopherson, 
Gaoli Chen, Sankha Chakrabarty and Yaqi Han for useful discussions, and 
Mark Hertzberg for pointing out an error in our original version of this 
paper \cite{MH}.  This work was supported in part by the U.S. Department 
of Energy under grant No. DE-FG02-97ER41029, and by the Heising-Simons 
Foundation under grant No. 2015-109.

\end{document}